\documentclass[sigconf]{acmart}

\usepackage[nolist]{acronym}
\usepackage{booktabs} % For formal tables
\usepackage{tikz}
\usetikzlibrary{positioning}
\usepackage{placeins }
\usepackage{subfig}

\usepackage{pdfcomment}

\renewcommand\footnotetextcopyrightpermission[1]{} % removes footnote with conference information in first column
\usepackage{multirow}
\usepackage{multicol}
\usepackage{amsmath}
\usepackage{array}
\usepackage{fullpage}
\usepackage{graphicx}
\usepackage{threeparttable}
\usepackage{wasysym}
\usepackage{xparse}
\usepackage{colortbl}
\usepackage{xspace}
\usepackage{boldline}
\usepackage{makecell}
\usepackage{tabularx}
\newcolumntype{Y}{>{\centering\arraybackslash}X}

\usepackage[binary-units=true]{siunitx}
\begin{document}

\title{Simulating Hybrid Aerial- and Ground-based Vehicular Networks with ns-3 and LIMoSim}

%\ifdefined\accepted
	\newcommand{\cni}{\affiliation{%
		\institution{Communication Networks Institute}
		%\streetaddress{Otto-Hahn-Straße 6}
		\city{TU Dortmund University}
		\state{Germany}
		\postcode{44227}\
	}}

	\author{Benjamin Sliwa, Manuel Patchou, Karsten Heimann, Christian Wietfeld}
	\cni
	\email{firstname.lastname@tu-dortmund.de}


%\fi

\begin{abstract}
	
%
% Introduction and Problem statement
%
Integrating \acp{UAV} into future \acp{ITS} allows to exploit their unique mobility potentials for improving the performance of services such as near-field parcel delivery, dynamic network provisioning, and aerial sensing.
In order to provide a controllable environment for the methodological performance analysis,  simulation frameworks need to support ground- and aerial-based mobility as well as the involved communication technologies.
%
% Solution appraoch
%
In this paper, we present the open source \ac{LIMoSim} framework for simulating hybrid vehicular networks within \ac{ns-3}. \ac{LIMoSim} implements a shared codebase coupling approach which integrates all required components in a single simulation process. The low-level mobility behaviors rely on well-known analytical models.
%
% Results
%
Different case studies discussing cutting-edge communication technologies such as \ac{C-V2X} and \ac{mmWave} are presented in order to illustrate how the proposed framework can be integrated into \ac{ns-3}-based network simulation setups.

\end{abstract}



\maketitle

\newcommand\single{1\textwidth}
\newcommand\double{.48\textwidth}
\newcommand\triple{.32\textwidth}
\newcommand\quarter{.24\textwidth}
\newcommand\singleC{1\columnwidth}
\newcommand\doubleC{.475\columnwidth}

\newcommand{\figurePadding}{0pt}
\newcommand{\figureTopPadding}{\figurePadding}
\newcommand{\figureBottomPadding}{\figurePadding}

\newcommand{\lte}{\ac{LTE}\xspace}
\newcommand{\lena}{\ac{LENA}\xspace}
\newcommand{\ltesim}{LTE-Sim\xspace}
\newcommand{\opnet}{Riverbed\xspace}
\newcommand{\simulte}{SimuLTE\xspace}
\newcommand{\omnet}{\ac{OMNeT++}\xspace}
\newcommand{\inet}{INET\xspace}
\newcommand{\itu}{ITU-R M.2135\xspace}

\newcommand{\net}{\ac{ns-3}\xspace}
\newcommand{\limosim}{\ac{LIMoSim}\xspace}
\newcommand{\mmWave}{\ac{mmWave}\xspace}

\newcommand{\intro}{}
\newcommand{\implementation}{\textbf{Implementation:}\xspace}
\newcommand{\results}{\textbf{Results:}\xspace}

\newcommand\tikzFig[2]
{
	\begin{tikzpicture}
		\node[draw,minimum height=#2,minimum width=\columnwidth,text width=\columnwidth,pos=0.5]{\LARGE #1};
	\end{tikzpicture}
}

\newcommand{\dummy}[3]
{
	\begin{figure}[b!]  
		\begin{tikzpicture}
		\node[draw,minimum height=6cm,minimum width=\columnwidth,text width=\columnwidth,pos=0.5]{\LARGE #1};
		\end{tikzpicture}
		\caption{#2}
		\label{#3}
	\end{figure}
}

\newcommand\pos{h!tb}

\newcommand{\basicFig}[7]
{
	\begin{figure}[#1]  	
		\vspace{#6}
		\centering		  
		\includegraphics[width=#7\columnwidth]{#2}
		\caption{#3}
		\label{#4}
		\vspace{#5}	
	\end{figure}
}
\newcommand{\fig}[4]{\basicFig{#1}{#2}{#3}{#4}{0cm}{0cm}{1}}

\newcommand\sFig[2]{\begin{subfigure}{#2}\includegraphics[width=\textwidth]{#1}\caption{}\end{subfigure}}
\newcommand\vs{\vspace{-0.3cm}}
\newcommand\vsF{\vspace{-0.4cm}}

\newcommand{\subfig}[3]
{
	\subfloat[#3]
	{
		\includegraphics[width=#2\textwidth]{#1}
	}
	\hfill
}

\newcommand\circled[1] % caution with using in captions: \protect \circled
{
	\tikz[baseline=(char.base)]
	{
		\node[shape=circle,draw,inner sep=1pt] (char) {#1};
	}\xspace
}
\begin{acronym}
	\acro{LENA}{LTE-EPC Network Simulator}
	\acro{HLA}{High Level Architecture}	
	\acro{DDNS}{Data-driven Network Simulation}	
	\acro{LTE}{Long Term Evolution}	
	\acro{V2X}{Vehicle-to-Everything}
	\acro{V2V}{Vehicle-to-Vehicle}
	\acro{C-V2X}{Cellular Vehicle-to-Everything}
	\acro{UE}{User Equipment}
	\acro{eNB}{evolved Node B}
	\acro{RSRP}{Reference Signal Received Power}
	\acro{CNI}{Communication Networks Institute}
	\acro{GCS}{Ground Control Station}
	\acro{CAM}{Cooperative Awareness Message}
	\acro{IPC}{Interprocess Communication}
	\acro{TCP}{Transmission Control Protocol}
	\acro{SUMO}{Simulation of Urban Mobility}
	\acro{TraCI}{Traffic Control Interface}
	\acro{UAV}{Unmanned Aerial Vehicle}
	\acro{UAS}{Unmanned Aerial System}
	\acro{ITS}{Intelligent Transportation System}
	\acro{ns-3}{Network Simulator 3}
	\acro{Veins}{Vehicles in Network Simulation}
	\acro{OMNeT++}{Objective Modular Network Testbed in C++}
	\acro{LIMoSim}{Lightweight ICT-centric Mobility Simulation}
	\acro{OSM}{OpenStreetMap}
	\acro{SPS}{Semi-persistent Scheduling}
	\acro{IDM}{Intelligent Driver Model}
	\acro{MIP}{Mixed Integer Programming}
	\acro{PDR}{Packet Delivery Ratio}
	\acro{VANET}{Vehicular Ad-hoc Networks}
	\acro{CUSCUS}{CommUnicationS-Control distribUted Simulator}
	\acro{AVENS}{Aerial VEhicle Network Simulator}
	\acro{FSTSP}{Flying Sidekick Travelling Salesman Problem}
	\acro{TSP}{Travelling Salesman Problem}
	\acro{TSP-D}{Travelling Salesman Problem with Drone}
	\acro{VRP}{Vehicle Routing Problem}
	\acro{PDP}{Pickup and Delivery}
	\acro{3GPP}{Third-Generation Partnership Project}
	\acro{DSRC}{Dedicated Short-range Communication}
	\acro{LOS}{Line-of-sight}
	\acro{NLOS}{Non-line-of-sight}
	\acro{mmWave}{millimeter Wave}
	\acro{UI}{User Interface}
	\acro{OpenGL}{Open Graphics Library}
	\acro{FL-AIR}{Framework libre AIR}
	\acro{RAT}{Radio Access Technology}
	\acro{WAVE}{Wireless Access for Vehicular Environment}
	\acro{CBR}{Constant Bitrate}
	\acro{UDP}{User Datagram Protocol}
\end{acronym}

\begin{tikzpicture}[remember picture, overlay]
\node[below=5mm of current page.north, text width=20cm,font=\sffamily\footnotesize,align=center] {Accepted for presentation in: Proceedings of the 2020 Workshop on Ns-3\vspace{0.3cm}\\\pdfcomment[color=yellow,icon=Note]{
@InProceedings\{Sliwa2020simulating,\\
	Author = \{Benjamin Sliwa and Manuel Patchou and Karsten Heimann and Christian Wietfeld\},\\
	Title = \{Simulating hybrid aerial- and ground-based vehicular networks with ns-3 and LIMoSim\},\\
	Booktitle = \{Proceedings of the 2020 Workshop on Ns-3\},\\
	Year = \{2020\},\\
	Address = \{Gaithersburg, Maryland, USA\},\\
	Month = \{Jun\},\\
\}
}};
\end{tikzpicture}

\section{Introduction}

%
% Integration of UAVs into ITSs
%
The integration of \acp{UAV} into smart city-based \acp{ITS}  \cite{Menouar/etal/2017a} will allow to exploit the third physical dimension in order to overcome the limitations of purely ground-based traffic systems \cite{Sliwa/etal/2019b}.
%
% Applications
%
Novel applications for hybrid vehicular networks such as drone-enabled parcel pickup and delivery \cite{Patchou/etal/2019a}, dynamic aerial-based network provisioning \cite{Cheng/etal/2018a} as well as aerial sensing \cite{Elloumi/etal/2018a} have been demonstrated in first feasibility studies.
%
% Communication
%
For the further development of these novel systems, the availability of reliable and efficient communication technologies is a basic requirement. This fact manifests in ongoing standardization initiatives such as 3GPP TR 36.777 \cite{3GPP/2017a} which aim to investigate the requirements for integrating aerial vehicles into cellular networks. The ongoing developments show that there is a \emph{convergence of mobility and communication}.

%
% Anticipatory communication
%
\emph{Anticipatory communication} \cite{Bui/etal/2017a} has emerged as a novel paradigm for wireless communication systems which aims to actively exploit measurable context information in order to improve decision processes such as data transmission scheduling \cite{Sliwa/etal/2019d}, routing, and handover. 
%
% Mobile robotic networs -> Controlled mobility
% 
\ac{UAV} networks -- as a sub-category of mobile robotic networks -- implement a form of \emph{controlled mobility}. Hereby, control routines are applied to execute a certain desired behavior (e.g., hovering over a centroid of ground users). Since knowledge about the planned mobility is inherently present within the mobile agents, those networks form a perfect match with anticipatory mobile networking mechanisms for proactive system optimization \cite{Sliwa/etal/2016a}.

%
% Fig. Scenario + Challenges + Research Tasks
%
\fig{}{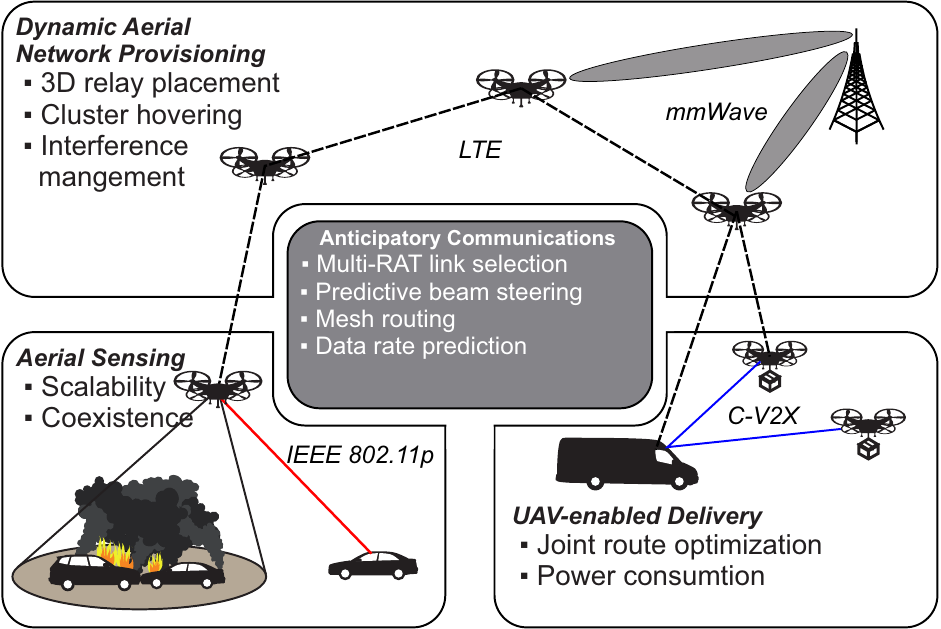}{Overview of applications, challenges, and communication technologies for hybrid vehicular networks.}{fig:scenario}
An overview of different applications for hybrid vehicular networks as well as challenges and research topics for the communication systems is illustrated in Fig.~\ref{fig:scenario}.

%
% Solution approach vs state of the art
%
The proposed simulation framework \ac{LIMoSim} aims to bring together ground- and aerial vehicular systems with anticipatory mobile networking. In previous work, we have presented an initial feasibility study for co-simulating ground-based and aerial vehicles \cite{Sliwa/etal/2019c}. In this paper, we focus on describing the interplay of \limosim and \net.
In particular, \limosim\footnote{The source code is available at \url{https://github.com/BenSliwa/LIMoSim_ns3}} makes the following contributions to the \net ecosystem:
%
% Contributions
%
\begin{itemize}
	\item An \textbf{integrated} approach for joint simulation of hybrid ground-based an aerial communication networks based on well-known analytical mobility models.
	\item Focus on \textbf{anticipatory} mobile networking through native integration of enablers for prediction models (e.g., mobility prediction, network quality maps).
	\item Online \textbf{3D visualization} based on \ac{OpenGL}.
\end{itemize}
%
%
%

%
% Structure
%
The remainder of this paper is structured as follows. After discussing related research work in Sec.~\ref{sec:related_work}, we provide an overview about the \limosim framework in Sec.~\ref{sec:approach} and discuss its integration into \net in details. Afterwards, the considered research methodology and the application of \limosim is shown in different case studies in Sec.~\ref{sec:results}.
\section{Related Work} \label{sec:related_work}

%
% COMMUNICATION
%
\textbf{Aerial and ground-based networks}:
A comprehensive summary that approaches a large variety of recent and future research topics related to \ac{UAV} communications is provided by Zeng et al in \cite{Zeng/etal/2019a}. While the existence of a dominant \ac{LOS} link is often a valid assumption for \emph{air-to-air} links, \emph{air-to-ground} communication is massively impacted by the dynamics between \ac{LOS} and \ac{NLOS} situations related to obstacle shadowing. Effects of the terrain profiles are further investigated by \cite{Hadiwardoyo/etal/2019a}.
Zhou et al. introduce an architecture model for enabling cooperative vehicular networking between cars and \acp{UAV} in \cite{Zhou/etal/2015a}.
%
% Technologies
%
A variety of communication technologies is applied for interconnecting the different vehicle types. While IEEE 802.11-based multi-hop networks \cite{Zhao/etal/2019a} have been in the research focus for several years, the integration of aerial vehicles into cellular communication networks is now actively being discussed \cite{3GPP/2017a}.

%
% Choi/etal/2016a - DSRC for mmWave beam steering
%
Apart from only using single technologies for interconnecting the different vehicles, multi-\ac{RAT} optimization \cite{Sepulcre/Gozalvez/2019a} has become an emerging research field. 
Choi et al. \cite{Choi/etal/2016a} propose a \ac{DSRC}-based exchange of position information in order to improve reduce the overhead of dynamic beam alignment for vehicular \ac{mmWave} networks.

%
% Cooperative routing
%
Due to the different mobility characteristics and the inherent resource constraints for aerial vehicles, many applications develop novel approaches for joint mobility optimization and cooperative routing \cite{Liu/etal/2019a}.
Two-echelon vehicle trajectory optimization methods for battery usage improvement are analyzed by the authors of \cite{Zeng/Zhang/2017a}.
Shang et al. \cite{Shang/etal/2019a} use the flexibility of the \ac{UAV} mobility to enhance the physical layer security for \ac{V2X} communications. In order to avoid eavesdropping, data transmissions between ground vehicles are forwarded by an intermediate aerial relay that establishes a virtually unobstructed \ac{LOS} between the users.
A communication-aware mobility model for \ac{UAV}-supported \ac{V2X} is proposed by \cite{Hadiwardoyo/etal/2019b}. Based on an attraction model, the \ac{UAV} automatically approaches the car which the lowest measured signal quality in order to avoid link loss within the served cluster of ground users.

For the case studies presented in Sec.~\ref{sec:results}, we apply existing \net implementations for \ac{mmWave} extension \cite{Ford/etal/2016a, Mezzavilla/etal/2018a}, \ac{C-V2X} mode 4 \cite{Eckermann/etal/2019a} and \ac{WAVE}-based IEEE 802.11p.

%
% SIMULATION FRAMEWORKS
%
\textbf{System-level network simulation} is the dominant performance analysis method for mobile and vehicular communication networks with the wireless research community \cite{Cavalcanti/etal/2018a}.
While ground-based mobility simulation -- often carried out with the \ac{SUMO} traffic simulator \cite{Lopez/etal/2018a} -- has already reached a highly mature state, \ac{UAV} mobility simulation is still in its infancy which has led to a variety of specialized simulation frameworks that target different use-cases and research topics.
%
% FlyNetSim
%
\emph{FlyNetSim} \cite{Baidya/etal/2018a} applies a middleware-based approach to couple \net with Ardupilot and focuses on hardware-in-the-loop simulations.
%
% CUSCUS
%
\emph{\ac{CUSCUS}} \cite{Zema/etal/2018a} provides a limited -- e.g., \ac{LTE} simulation is currently not supported -- interconnection of \net and \ac{FL-AIR} based on Linux containers. 
%
% OpenUAV
%
\emph{OpenUAV} \cite{Schmittle/etal/2018a} is an open source test bed for \ac{UAV} research featuring rich visualization capabilities and cloud-based simulation support. However, it focuses on individual mobility control and does not provide capabilities for simulating actual communication technologies.
%
% Corner-3D
%
\emph{Corner-3D} \cite{Ferlini/etal/2019a} focuses on providing a realistic representation of typical \ac{UAV} obstacle-related channel dynamics without actually simulating the \ac{UAV} mobility itself.
%
% Generic UAV mobility models in ns-3 -> Lacks a physical acceleration model of the UAV
%
Within the \ac{ns-3} ecosystem, generic random-based obstacle-aware \ac{UAV} mobility models have been introduced by the authors of \cite{Regis/etal/2016a}. However, these approaches focus on high-level mobility and lack of a realistic representation of the acceleration dynamics.
In comparison to these approaches, \ac{LIMoSim} combines high-level mobility modeling with validated low-level acceleration models. This method is comparable to the common approach used used in the car simulation domain.

%
% INTERCONNECTION
%
In order to interconnect the mobility simulation with a network simulator, the majority of existing approaches applies a \ac{HLA}-based method. Popular examples are \emph{iTETRIS} \cite{Rondinone/etal/2013a} which integrates \ac{SUMO} \cite{Lopez/etal/2018a} and \net via the \ac{TraCI} protocol. A similar workflow is implemented by \emph{\ac{Veins}} \cite{Sommer/etal/2011a} for \emph{\ac{OMNeT++}}.
%
% HLA
%
Although \ac{HLA} has a long tradition in the performance analysis of wireless communication networks, as it allows to derive highly accurate simulation setups based on specialized tools, it has a number of disadvantages:
%
% Disadvantages of IPC-based coupling
%
\begin{itemize}
	\item \textbf{Complexity}: Although hybrid vehicular networks can be simulated with a combination of existing tools, this approach is not very practical. It requires to execute and synchronize at least three different system processes (network simulator, car mobility simulator, \ac{UAV} mobility simulator) which results in a highly complex simulation setup.
	\item \textbf{Performance}: As a consequence of the setup complexity, computation and memory resources are wasted on the required coordination within the simulation setup itself. This aspect is further analyzed by the authors of \cite{Hu/etal/2017a} which analyze the scalability of integrated and \ac{HLA}-based co-simulation approaches.
	\item \textbf{Maintenance}: Since the different frameworks are developed further independently from each other, compatibility issues might occur when new framework versions are introduced.
	\item \textbf{Usability}: For anticipatory mobile networking, the protocol-based synchronization is a non-intuitive way of data exchange between the mobility and communication domains which requires dedicated serialization and parsing for each newly integrated method.
\end{itemize}
%
% LIMoSim
%
Instead, the integrated simulation approach of \limosim provides a more lightweight alternative as it brings together the different logical domains in a single system process. In addition, the shared-codebase coupling -- mobility and communication interact based on \texttt{C++} pointers -- explicitly targets the development of novel anticipatory communication methods that exploit synergies between the different logical domains.

\section{Simulating Hybrid Vehicular Networks with LIMoSim and ns-3} \label{sec:approach}

%
% Operation modes
%
Although the regular operation mode is the joint simulation of \limosim with a coupled \net instance, \limosim does not have any code dependencies to the latter or any other external library. Objects of the \limosim core are not aware of their \net execution environment. This design approach allows to execute \emph{standalone} simulations focusing only on the mobility behavior of the vehicles. 
%
% Fig. Architecture model
%
\fig{}{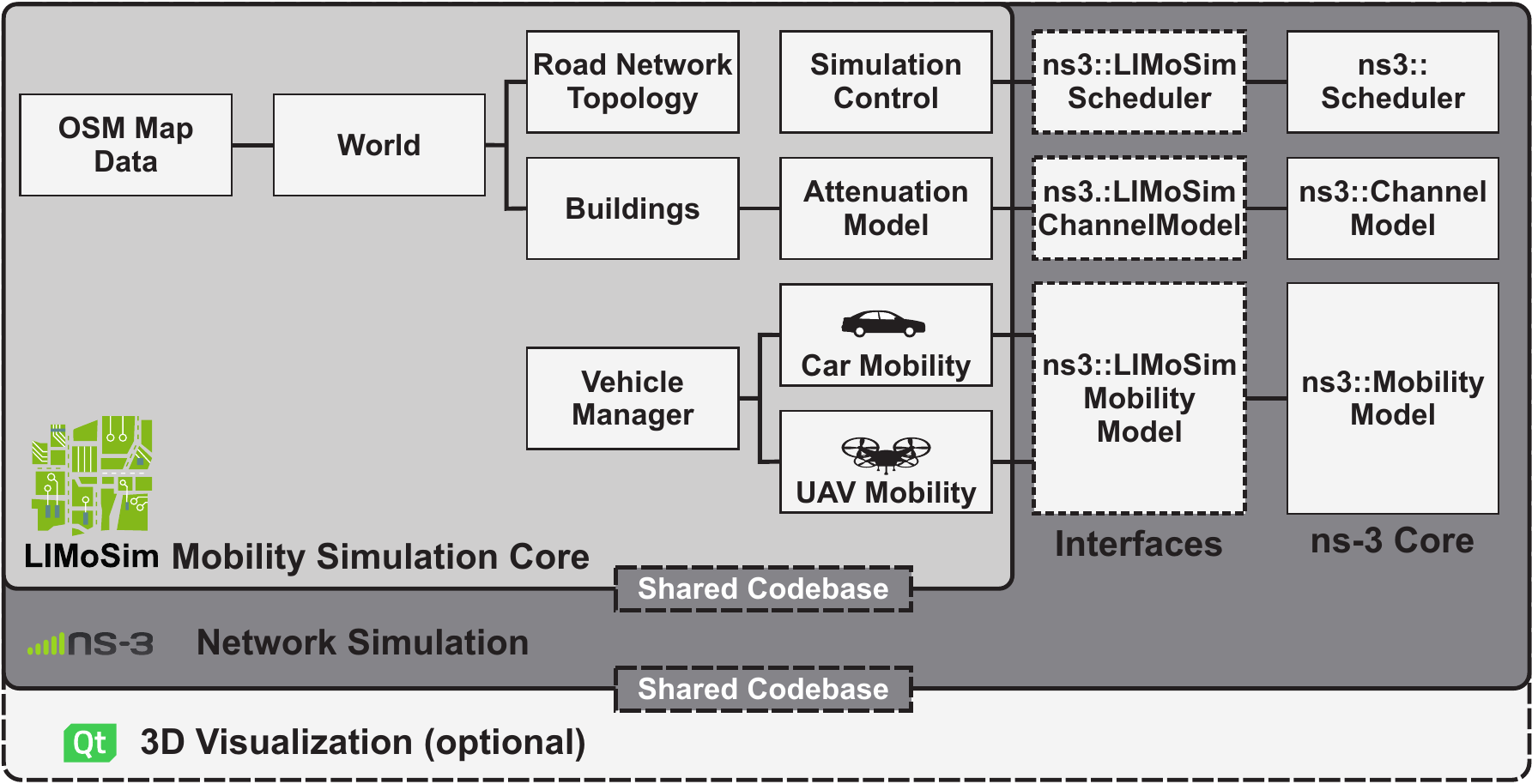}{Simplified class diagrams for the integration of \limosim into \net.}{fig:architecture}
The interplay between \limosim and \net as well as the most important modules is shown in the system architecture model in Fig.~\ref{fig:architecture}. 
%
% Visualization 
%
Optionally, the mobility behavior of the vehicles and their 3D environment is visualized online based in an \ac{OpenGL}-based \ac{UI} which is implemented in \texttt{Qt C++} (see Fig.~\ref{fig:map}).
%
% EPS Export
%
In addition to the online visualization capabilities, \limosim features a native rendering engine for exporting screenshots of the 3D environment in a vector graphics format.
In the following paragraphs, we give an overview about simulation control, mobility handling and obstacle-aware channel modeling.

\subsection{Coupling of LIMoSim and ns-3}

While the vast majority of existing approach relies on \ac{IPC}-based coupling (see Sec.~\ref{sec:related_work}), \limosim implements a fundamentally different method to interconnect the logical domains. Instead, its mobility simulation core is directly embedded into \net using a \emph{shared codebase} coupling method. 
In order to support the explicit focus on the development of anticipatory vehicular communication systems, \limosim allows intuitive pointer-based direct interactions between \texttt{C++} objects of the two domains.
\limosim uses its own logical event handling system in order to be agnostic to and independent from the coupled network simulator. However, if \limosim is coupled to \net, the event scheduler of the latter takes over control about the event handling mechanisms. 
%
% Fig. Event synchronization
%
\fig{}{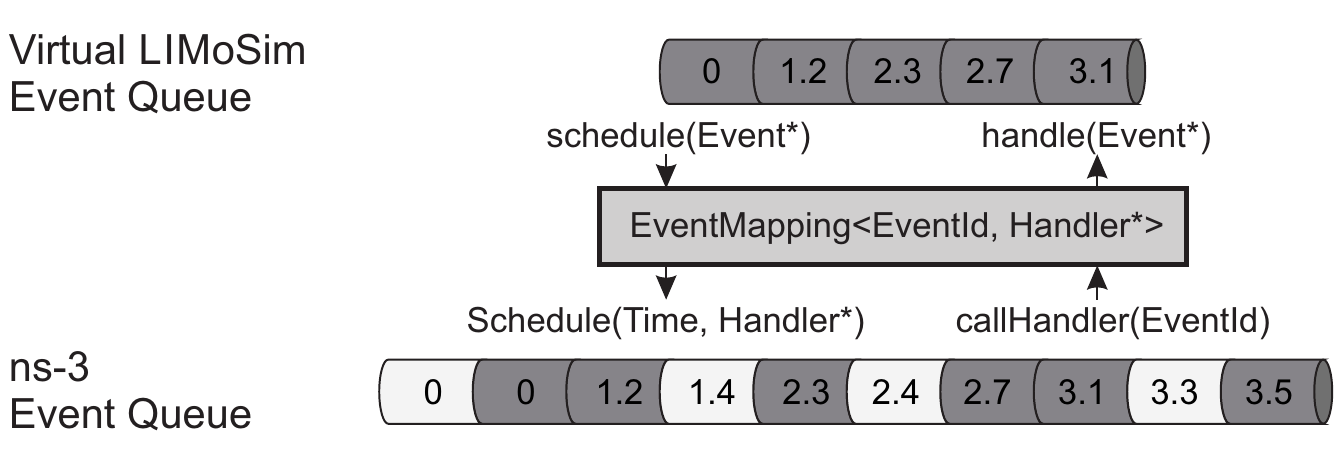}{Mechanism for the event synchronization between \limosim and \net.}{fig:event_sync}
Fig.~\ref{fig:event_sync} illustrates the involved event synchronization process. Mobility related events issued by \limosim are seamlessly integrated in the \net event queue and transparently handled through a mediator class which transforms the events between the different simulation domains and invokes the corresponding event handlers.

%
% Joint Synchronization Setup Flow
%
\fig{}{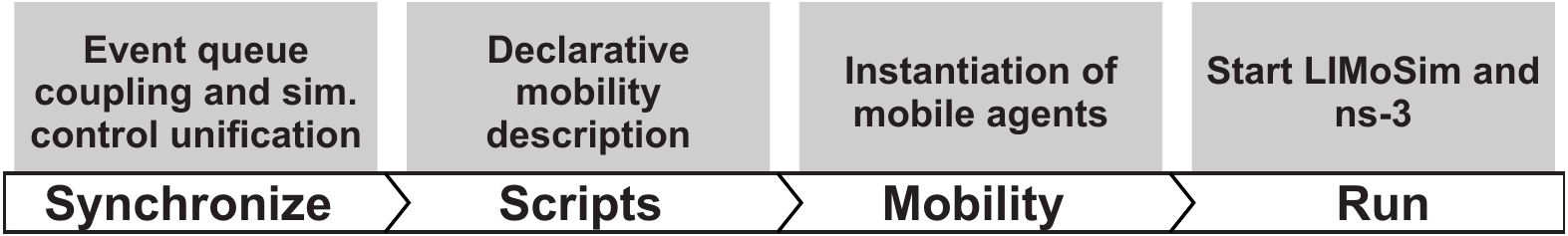}{Logical steps that establish a joint simulation setup.}{fig:flow}
For the establishment of a joint simulation setup, different steps are processed sequentially. Fig.~\ref{fig:flow} provides an overview about the resulting processing pipeline. The first step synchronizes both event queues and unifies the simulation control. This allows both frameworks to manipulate the event scheduling mechanisms of the active simulation -- e.g., for pausing the simulation with the \limosim \ac{UI}.
Next, the \net simulation script is processed. A dedicated helper extension is used to define \limosim mobility is associated to \net nodes in a declarative fashion.

The mobility definitions are then used in the next step by the helper extension to automatically instantiate the mobile agents in \limosim and configure them to be linked to their \net counterparts.
The linker installs an \texttt{ns3::LIMoSimMobilityModel} that is derived from the \net base class \texttt{ns3::MobilityModel} on the \net nodes. In addition, mobile agents which do not belong to the \limosim domain -- e.g., purely \net-based entities -- can be registered in \limosim for visualization purposes

\subsection{Hierarchical Mobility Modeling}

Within \net, agent-based vehicular mobility simulation is performed based on \texttt{ns3::LIMoSimMobilityModel} which acts as an interface between the two framework domains. It is derived as a subclass of \texttt{ns3::MobilityModel} class of \net and supports \texttt{ns3::MobilityHelper}-based simplified installation on \net nodes.
All \limosim vehicles are derived from the abstract \texttt{LIMoSim::Vehicle} class and are automatically registered to the event handling system upon instantiation. Further details about the analytical foundations of the vehicular mobility models are described in \cite{Sliwa/etal/2019c} and \cite{Sliwa/etal/2017b}. Fig.~\ref{fig:mobility} summarizes core components of the hierarchical mobility models for both agent types which are further described in the following paragraphs.

%
% GROUND-BASED MOBILITY
%
\textbf{Ground vehicle mobility} models are implemented within the \texttt{LIMoSim::Car} class. It consists of two main submodels:
%
% Car Mobility
%
\begin{itemize}
	\item High-level behaviors are represented by a \textbf{strategic model} which is responsible for target definition and routing processes and supports random as well as deterministic path planning methods.
	\item Cruise control and velocity dynamics are handled by a \textbf{follower model} which updates the current acceleration of a vehicle with respect to the velocities of nearby traffic participants and the traffic rules. For this purpose, the well-known	\ac{IDM} is implemented according to \cite{Treiber/Kesting/2013a}.
\end{itemize}
%
%
%

%
% AERIAL MOBILITY
%
\textbf{Aerial mobility} relies on the hierarchical model proposed by \cite{Reynolds/1999a} which consists of three logical layers that are brought together in the \texttt{LIMoSim::UAV} class:
\begin{itemize}
	%
	% Action Selection
	%
	\item \textbf{Action selection} specifies the general behavior characteristics of the \ac{UAV} and allows to implement \emph{role}-based primitives (e.g., \emph{aerial sensors} aim to stay close to defined ground vehicles, \emph{aerial relays} maintain \ac{LOS} to a cellular base station).
	
	%
	% Steerings
	%
	\item \textbf{Steerings} are high-level mobility routines for a well-defined task that are executed in parallel. They are used for following a defined trajectory, for avoiding collisions with buildings and other vehicles, and for maintenance of a swarm coherence which ensures a certain level of connectivity. Within each update iteration, the result of each steering is a \emph{steering} vector which represents the desired movement in the next step. The final steering vector is computed as a weighted average of all individual vectors.
		
	%
	% Locomotion
	%
	\item \textbf{Locomotion} represents the physical motion and separates the logical vehicle control from the actual execution platform. Within \limosim, these low-level mobility functions based on analytical 3D acceleration and orientation models according to \cite{Luukkonen/2011a}.
	%
	% Low-level mobility and power consumption
	%
	On this layer, also the propulsion-related power consumption is computed based on the model of \cite{Yacef/etal/2017a} which allows to simulate joint optimization of mobility and communication for battery lifetime improvements.
\end{itemize}
%
%
%

%
% MOBILITY PREDICTION
%
\textbf{Mobility prediction} is an enabling method for anticipatory mobile networking. Thus, \limosim provides a mobility control-aware prediction mechanism that allows to forecast the future position $\mathbf{P}(t+\tau)$ for a given prediction horizon $\tau$.
The default implementation for the \ac{UAV} mobility prediction is based on the proposed hierarchical prediction model of \cite{Sliwa/etal/2016a} which exploits knowledge about steering vectors as well as waypoint information if available. A similar method is implemented for the ground-based vehicles where navigation system knowledge is used to forecast position estimates in a trajectory-aware manner. The effectiveness of this approach has been proven in real world experiments \cite{Sliwa/etal/2019d} where mobility predictions are jointly used with network quality maps in order to schedule vehicular sensor data transmissions with respect to the expected network quality.
All prediction mechanisms are impacted by uncertainties in the actual low-level mobility dynamics which depend on the traffic dynamics.

%
% Fig. Mobility 
%
\fig{}{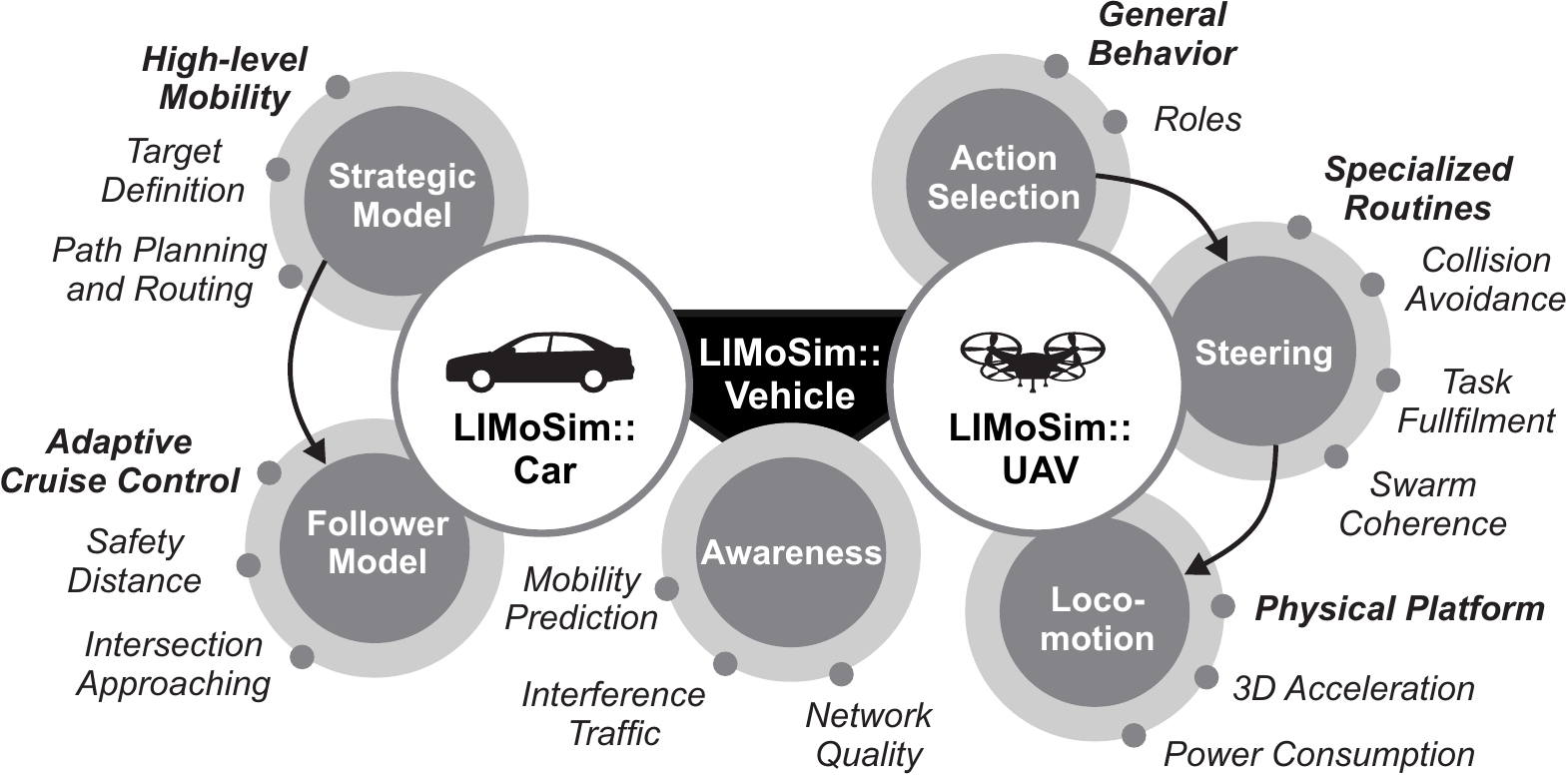}{Overview about the core components of the hierarchical mobility models for cars and \acp{UAV}.}{fig:mobility}

\subsection{Air-to-Ground Channel Modeling}

%
% OSM
%
\limosim provides native support for \ac{OSM} data and can optionally represent buildings as three-di\-men\-sional obstacles that cause attenuation to the radio signals via the \texttt{LimoSim::Building} class.
%
% Buildings as physical obstacles
%
Although \net itself provides capabilities for simulating shadowing-related attenuations, we decided to implement this feature in the \limosim domain since buildings act as physical obstacles that require collision avoidance routines for the aerial vehicles and are logically non-communicating entities of the environment. \texttt{C++}-level access to buildings and all other world objects (e.g., road segments) is provided via the \texttt{LimoSim::World} singleton.
For given receiver and transmitter positions $\mathbf{P}_{\text{RX}}(t)$ and $\mathbf{P}_{\text{TX}}(t)$, the attenuation model computes the three-dimensional obstructed distance $d_{\text{obs}}$ with respect to the intermediate building intersections.  
%
% UAV channel characteristics: LOS vs NLOS
%
Within \net, this information is then utilized for channel modeling with the \texttt{ns3::LIMoSimChannelModel}. In particular, the typical air-to-ground channel dynamics (see Sec.~\ref{sec:related_work}) between \ac{LOS} and \ac{NLOS} situations can be modeled automatically.
%
% Cache
%
A caching strategy is used to allow a resource efficient usage of the \texttt{ns3::LIMoSimChannelModel} to determine attenuation caused by the buildings in the simulation scenario. Within the simulations, the channel conditions are frequently re-evaluated by \net which leads to path loss computations being repeated multiple times for the same or similar spatial configurations, thus yielding the same results. Caching allows to reuse the results of previous computations for identical or similar receiver and transmitter positions in order to reduce the computational overhead.

\section{Case Studies} \label{sec:results}

In this section, we present two case studies that show the usage of \limosim in hybrid mobility applications and in coexistence with established \net extension frameworks. The traffic patterns of the communicating vehicles are chosen with respect to the application-specific requirements discussed in \cite{Zeng/etal/2019a}.
%
% Fig. Screenshot
%
\fig{}{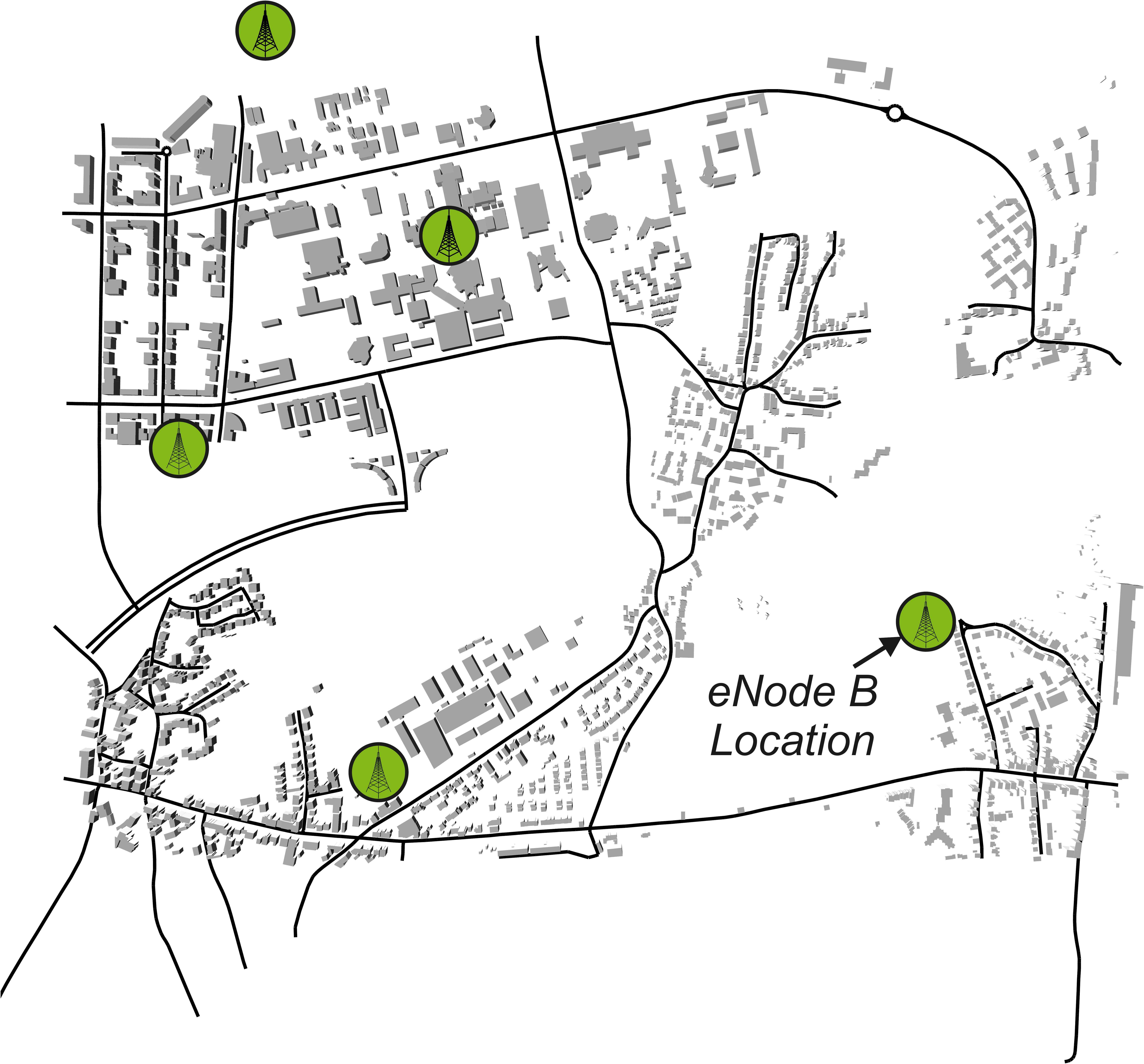}{Overview about the considered evaluation scenario. (Map data: $\copyright$ OpenStreetMap contributors, CC BY-SA).}{fig:map}
The evaluations are performed within a suburban environment near a university campus. Fig.~\ref{fig:map} shows a map of the resulting simulation scenario within \limosim. The simulation parameters are summarized in Tab.~\ref{tab:parameters}. 

%
% Tab. Parameters
%
\newcommand{\sideHeader}[3]
{
	\multirow{#1}{*}{
		\rotatebox[origin=c]{90}{
			\parbox{#2}{\centering \textbf{#3}}
		}
	}
}

\begin{table}[ht]
	\small
	\centering
	\caption{Simulation Parameters}
	\begin{tabularx}{1\columnwidth}{p{0.1cm}p{3.1cm}p{3.8cm}}
		\toprule
		
		\sideHeader{3}{2.3cm}{General} 
		& \textbf{Parameter} & \textbf{Value} \\

		\midrule
		& Simulation duration per run & \SI{30}{\minute} \\
		& Number of simulation runs & 50 \\
		& UAV fleight height & \SI{30}{\meter} \\
		& Channel model & ns3::LIMoSimChannelModel \\

		\midrule
		\sideHeader{3}{1cm}{\ac{C-V2X}} 
		& Carrier frequency & \SI{5.9}{\giga\hertz} \\
		& Bandwidth & \SI{20}{\mega\hertz} \\
		& $P_{\text{TX}}$ (\acs{UE}) & \SI{23}{\decibel m} \\

		\midrule
		\sideHeader{3}{1.4cm}{\ac{LTE}} 
		& Carrier frequency & \SI{2.1}{\giga\hertz} \\
		& Bandwidth & \SI{20}{\mega\hertz} \\
		& $P_{\text{TX}}$(\acs{UE}) & \SI{23}{\decibel m} \\
		& $P_{\text{TX}}$(\acs{eNB}) & \SI{43}{\decibel m} \\
		
		\midrule
		\sideHeader{3}{1.7cm}{\ac{mmWave}} 
		& Channel model & 3GPP UMi Street Canyon (LOS)\\
		& Carrier frequency & \SI{28}{\giga\hertz} \\
		& Antenna array & Planar eight--by--eight \\
		& Beamforming & Analog \\
		& Beam alignment & Geometry--based LOS (ideal) \\
		
		\bottomrule
		
	\end{tabularx}
	\label{tab:parameters}
\end{table}

\subsection{Case Study 1: \acp{UAV} as Aerial Sensors}

%
% INTRODUCTION: Aerial sensors in smart cities
%
\intro In the first case study, \acp{UAV} are exploited as \emph{aerial sensor} nodes that provide ground vehicles with potentially safety-relevant information for raising their situation-awareness. Similar to the ongoing discussions in \ac{V2V} networking, the usage of different communication technologies are in the focus of the analysis.
%
%
%

%
% IMPLEMENTATION
%
\implementation Within \limosim, the use case is modeled with five vehicle pairs composed of one \ac{UAV} and one car each. The latter move freely through the whole scenario based on a random direction mobility model. All \acp{UAV} operate at a constant flying height and aim to maintain a close distance to their assigned ground vehicle. Each 100~ms, the \acp{UAV} transmit \acp{CAM} consisting of 190~Byte sensor data to their corresponding cars. 
%
% Communication
%
Three different communication technologies are applied and compared using the same mobility configuration. \ac{LTE} (based on  \ac{LENA} \cite{Baldo/etal/2011a}) is used as benchmark technology that implements a \emph{centralized} medium access approach. In addition, we compare \ac{WAVE}-based IEEE 802.11p and \ac{C-V2X} (based on \cite{Eckermann/etal/2019a}) as representatives of \emph{decentralized} medium access approaches. For \ac{LTE}, the \acp{eNB} are positioned according to corresponding real world locations as shown in Fig.~\ref{fig:map}.

%
% RESULTS
%
\results 
%
% Fig. PDR and delay of aerial sensors
%
\begin{figure}[]
	%\captionsetup[subfigure]{labelformat=empty}
	\centering
	\subfig{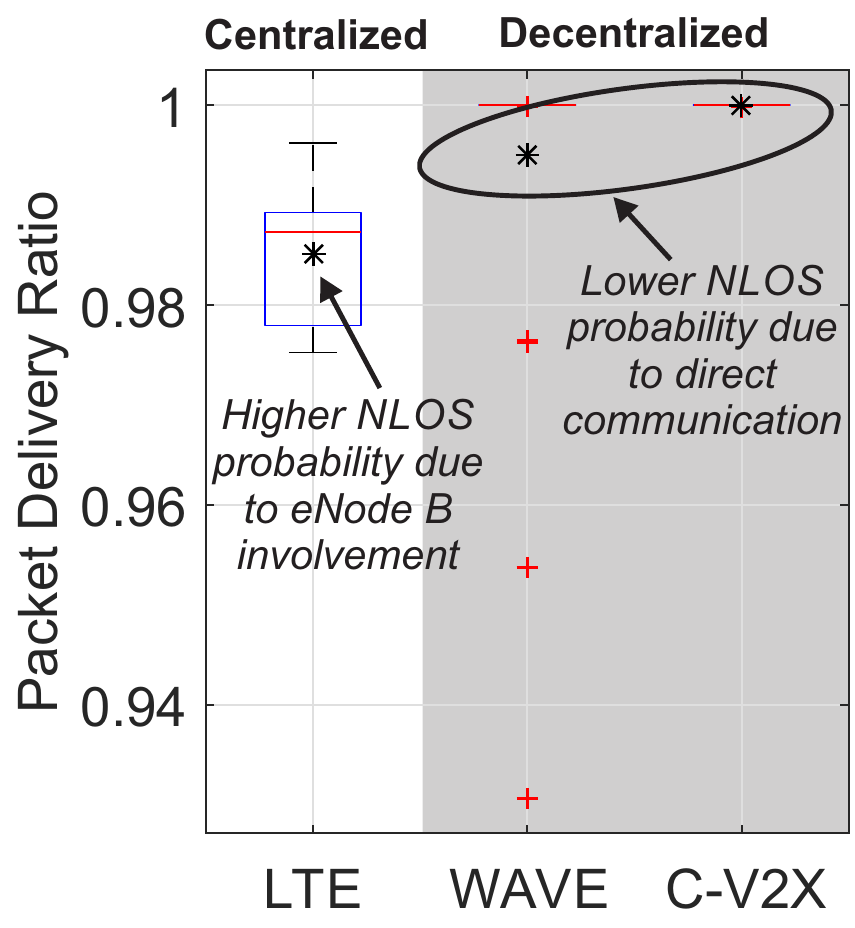}{0.22}{}
	\subfig{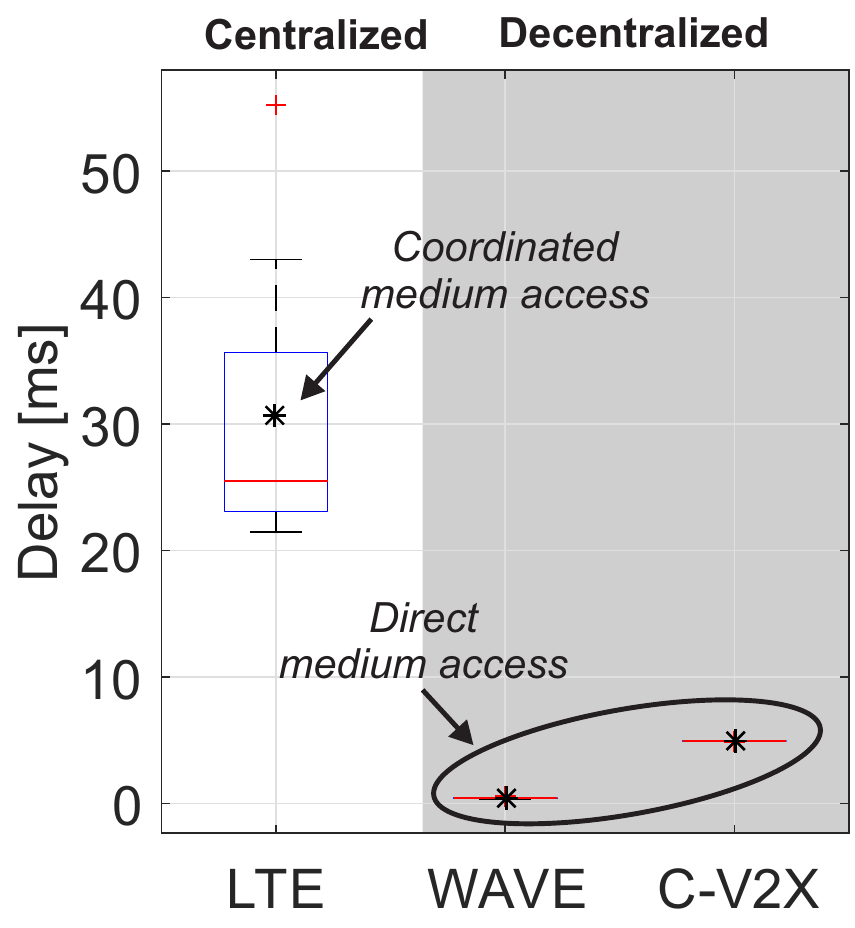}{0.22}{}
	
	\caption{Comparison of end-to-end performance metrics for different communication technologies in the aerial sensors use case.}
	\label{fig:aerial_sensors_box}
\end{figure}
The overall results for \ac{PDR} and delay for the different communication technologies are shown in Fig.~\ref{fig:aerial_sensors_box}.
%
% PDR
%
All technologies are able to provide robust communication links. However, the decentralized approaches represented by \ac{C-V2X} and \ac{WAVE} exhibit higher reliability with a \ac{PDR} very close to 1. Due to the direct transmission path between sender and receiver, the probability for \ac{NLOS} situations is lower than for \ac{LTE} where the \ac{eNB} is also involved in the communication process. In the considered scenario, \ac{C-V2X} achieves a slight better reliability than \ac{WAVE} due to the \ac{SPS}-based medium access which takes the previous resource reservations into account in order to avoid resource conflicts in the future resource reservation periods.
%
% Delay
%
A similar tendency between both medium access approaches is observed when considering the latency. The decentralized approaches -- which implement direct medium access strategies -- yield a smaller latency as opposed to \ac{LTE} which handles the resource reservations centrally.

% Fig. Indicator vs time
%
\fig{}{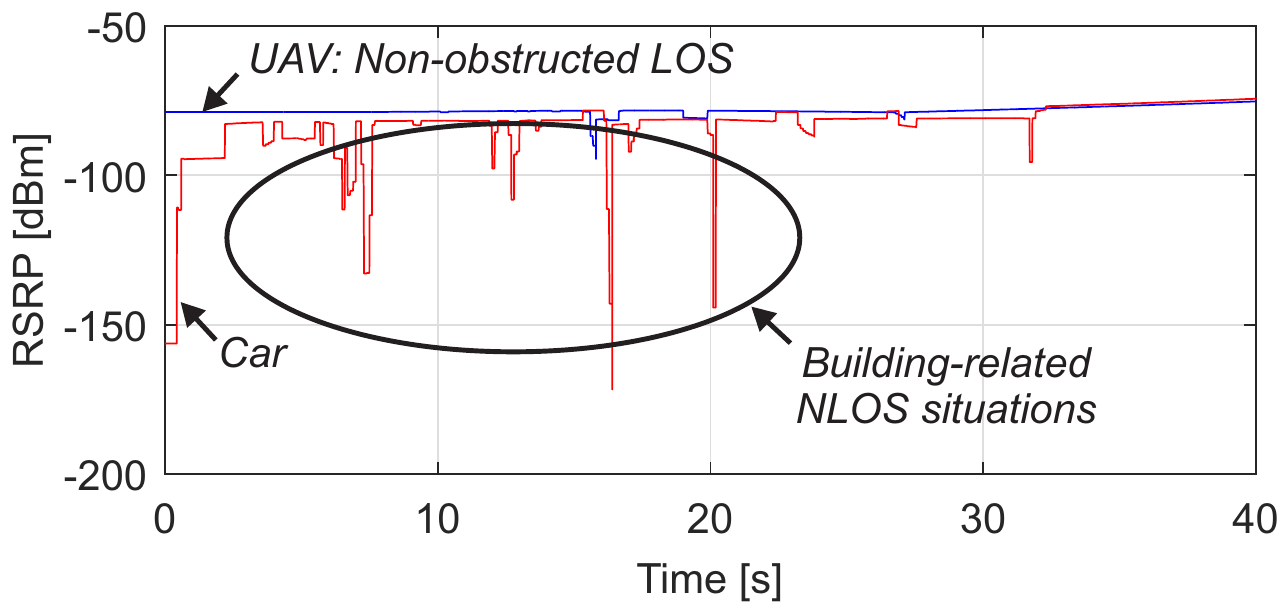}{Example temporal behavior of \acs{RSRP} for aerial and ground vehicles.}{fig:time_example}
For \ac{LTE}, an example for the temporal dynamics of the \ac{RSRP} for a cooperating pair of ground and flying vehicles is shown in Fig.~\ref{fig:time_example}. The different obstacle-related attenuation characteristics for the ground- and air-based vehicles can be clearly distinguished. While the car is subject to high attenuation peaks caused by signal shadowing from the nearby buildings, the flying altitude of the \ac{UAV} results in a non-obstructed \ac{LOS} to the \ac{eNB} for the whole shown time period.

%
% Fig. Acc vs time
%
\begin{figure}
	%\captionsetup[subfigure]{labelformat=empty}
	\centering
	
	\subfig{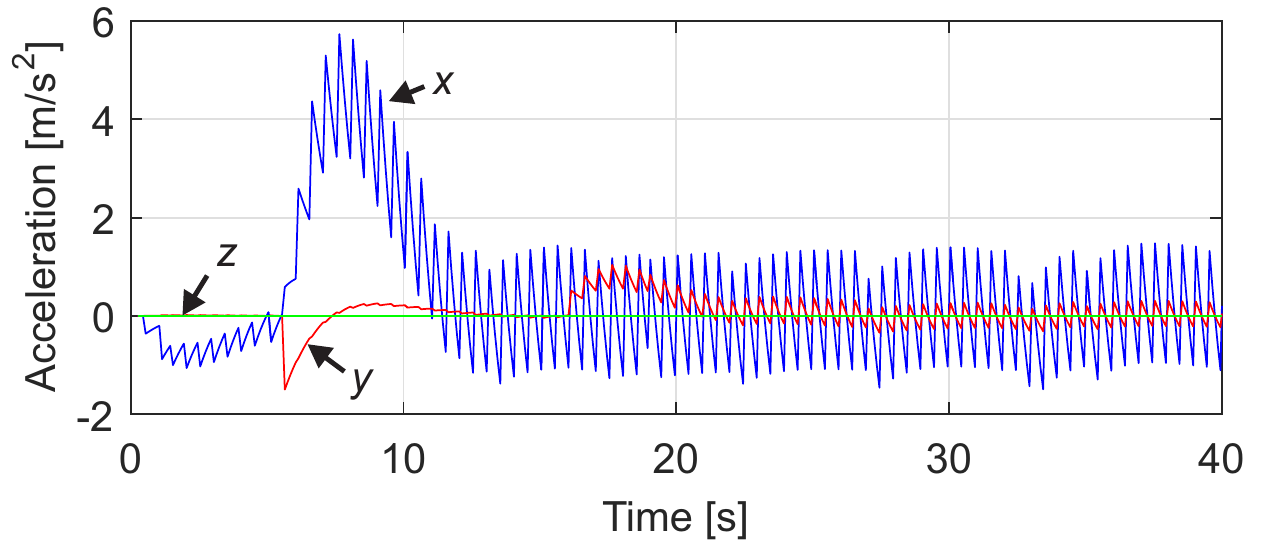}{0.45}{}
	\subfig{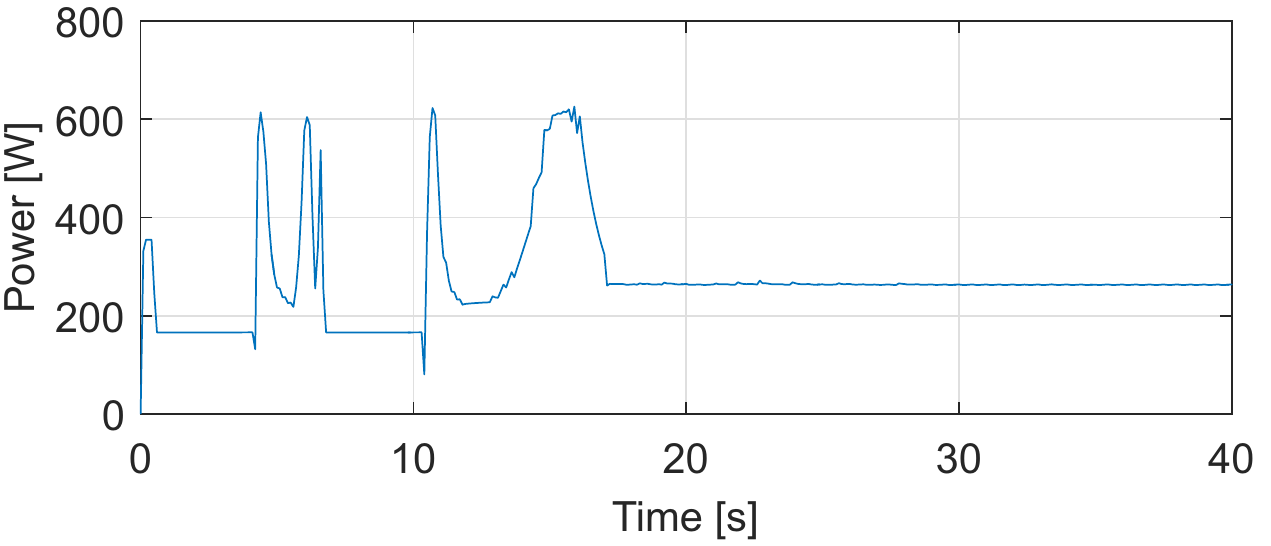}{0.45}{}
	
	\caption{Exemplary excerpt of the temporal dynamics of the three-dimensional acceleration and resulting power consumption for an example \ac{UAV} operating at a constant operation height.}
	\label{fig:mobility_time}
\end{figure}
Furthermore, an example for the interdependency between acceleration dynamics and resulting power consumption for the \ac{UAV} is shown in Fig.\ref{fig:mobility_time}. Since the \ac{UAV} operates at a constant flying height in the considered case study, there is no acceleration in the $z$ dimension. The resulting power consumption is the effect of the acceleration dynamics.

\subsection{Case Study 2: Millimeter Wave-based Data Transfer}

%
% INTRODUCTION
%
\intro Due to the vast amount of available radio spectrum, \mmWave communications is appealing for increasing throughput demands within 5G mobile radio networks and beyond. However, the higher frequencies offer more challenging radio conditions, which are believed to be compensated by means of beamforming, antenna arrays and directional communications, among others. This means, exploiting the antennas' electronically steerable, directional gains is crucial for a stable connectivity. Subsequently, the utilization of beamforming antennas necessitates a proper alignment of the beams. For this reason, the applicability of this approach within dynamic scenarios and user mobility states a major field of scientific research.
Together with the \net \mmWave module, \limosim offers the possibility of merging mobility simulation into \mmWave beam alignment related research topics. In a first step, the physical mobility characteristics are provided for exemplary channel models. However, more sophisticated models and communication--aware mobility may be implemented on demand by leveraging the prepared interfaces and subroutines.

%
% Fig. Beam pattern
%
\begin{figure}[] 
	\centering
	\includegraphics[width=\columnwidth]{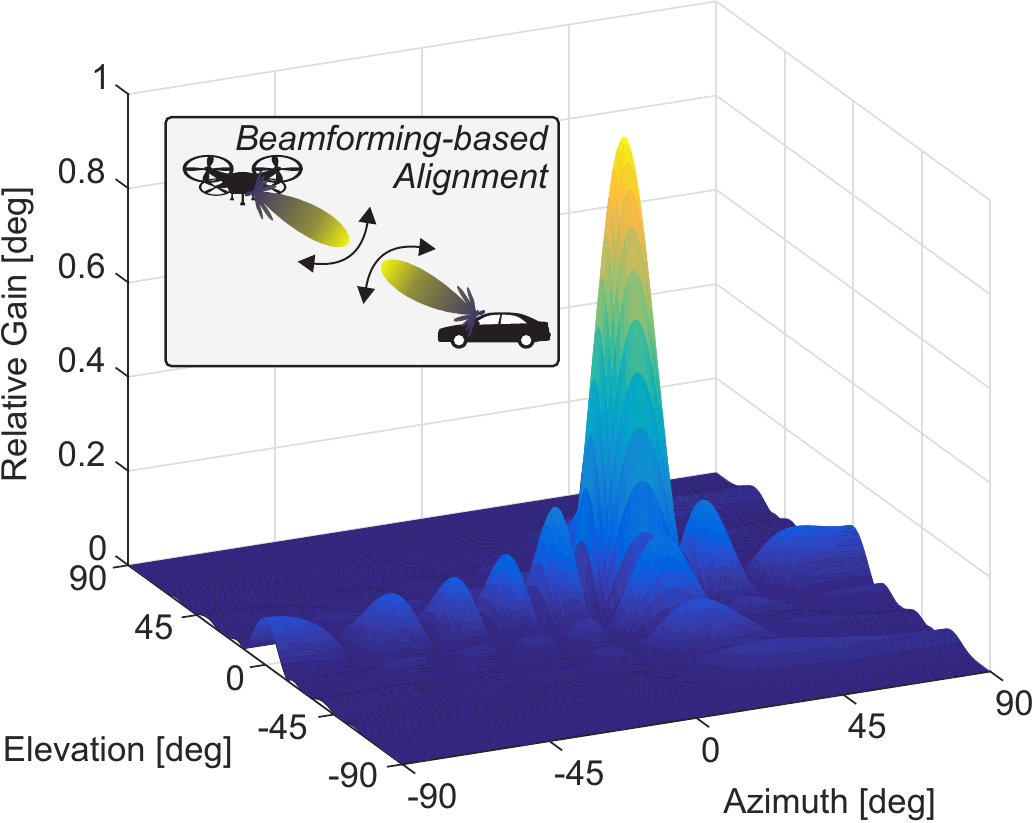}
	\caption{Exemplary pencil beam pattern for pointing direction of $\text{azimuth}=\SI{20}{\degree}$ and $\text{elevation} = \SI{0}{\degree}$ generated by means of analytical models of microstrip patch elements and an eight--by--eight boresight array with half--wavelength element spacing.}
	\label{fig:pencilbeam}
\end{figure}
%
%
%
%
% IMPLEMENTATION
%
\implementation For the assessment of the behavior of \mmWave-based data transfer in vehicular scenarios, a common analytical model of the pencil beam characteristic is implemented according to \cite{Balanis/2016a}. An eight--by--eight planar broadside array is assumed to contain microstrip patch elements at a half--wavelength spacing. Due to the patch characteristic, the angular coverage space is limited to reasonable pointing directions with a maximum deviation from boresight of \SI{60}{\degree}. In Fig.~\ref{fig:pencilbeam}, a model of the directional antenna gain is illustrated for an exemplary pointing direction of \SIlist{20;0}{\degree} for azimuth and elevation, respectively. This antenna model can be regarded as a generic implementation, which may be extended as required. However, it suffices for simulating beam alignment of transmitter and receiver.
The antenna gains derived using the analytical pencil beam characteristic from \cite{Balanis/2016a} are supplied to the \mmWave module \cite{Mezzavilla/etal/2018a} thus enabling online beamforming gain computation. An ideal beam tracking according to the geometric \ac{LOS} direction is used as preliminary beam alignment method for the \mmWave link. 
Within the case study, a \mmWave communication link between a base station \ac{UAV} and a ground vehicle as mobile subscriber is considered. The scenario defines a travel route for the ground vehicle, while the \ac{UAV} is following the latter at a constant altitude. Simultaneously, the \mmWave radio link is used for data streaming. We consider different intensities of \ac{UDP}-based \ac{CBR} traffic load and compare the behavior of the \mmWave data transfer with a reference \ac{LTE}-based setup.

%
% RESULTS
%
\results At first, the temporal dynamics of the \ac{CBR} data stream are analyzed for both technologies. Fig.~\ref{fig:throughput_time} shows an excerpt of the resulting behavior characteristics for a targeted traffic load of 65~MBit/s. 
%
% Fig. Throughput vs Time
%
\fig{}{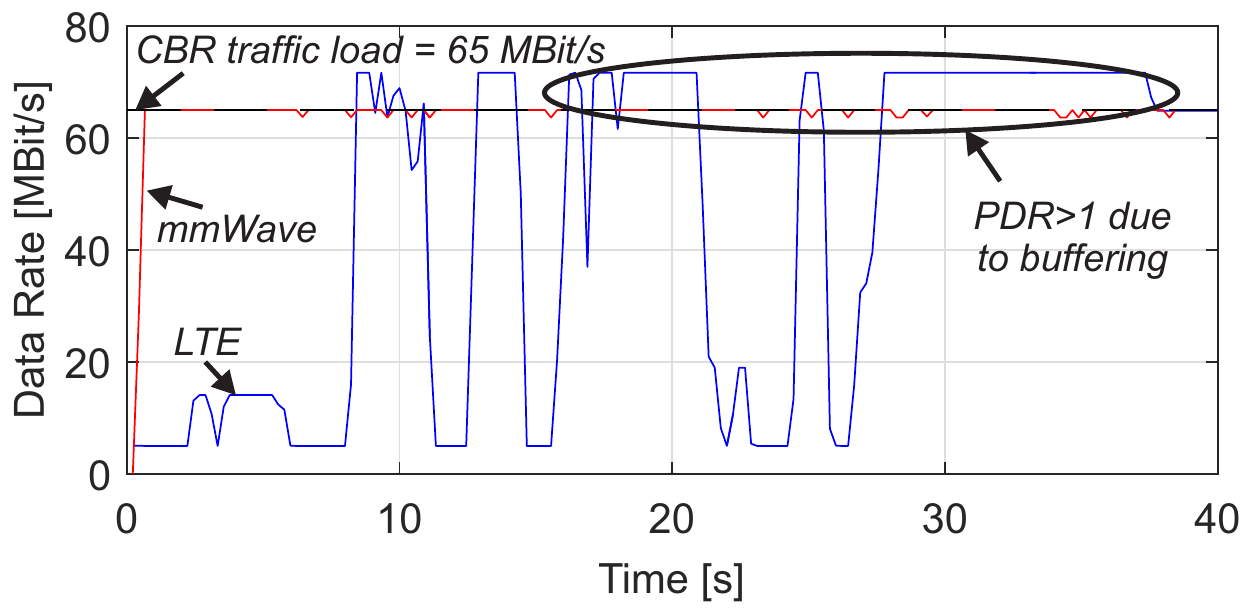}{Temporal behavior of the resulting data rate for LTE and mmWave with 65 MBits/s traffic load.}{fig:throughput_time}
While \mmWave provides a constant performance level close to the targeted traffic load, the \ac{LTE} transmission link behaves much more dynamic and alternates between periods of low and high data rates.
The latter is because the transmission buffers are filled and flushed with respect to the network congestion.
Hence, the resulting data rate sporadically exceeds the targeted traffic load.

\begin{figure}[]
	%\captionsetup[subfigure]{labelformat=empty}
	\centering
	
	\subfig{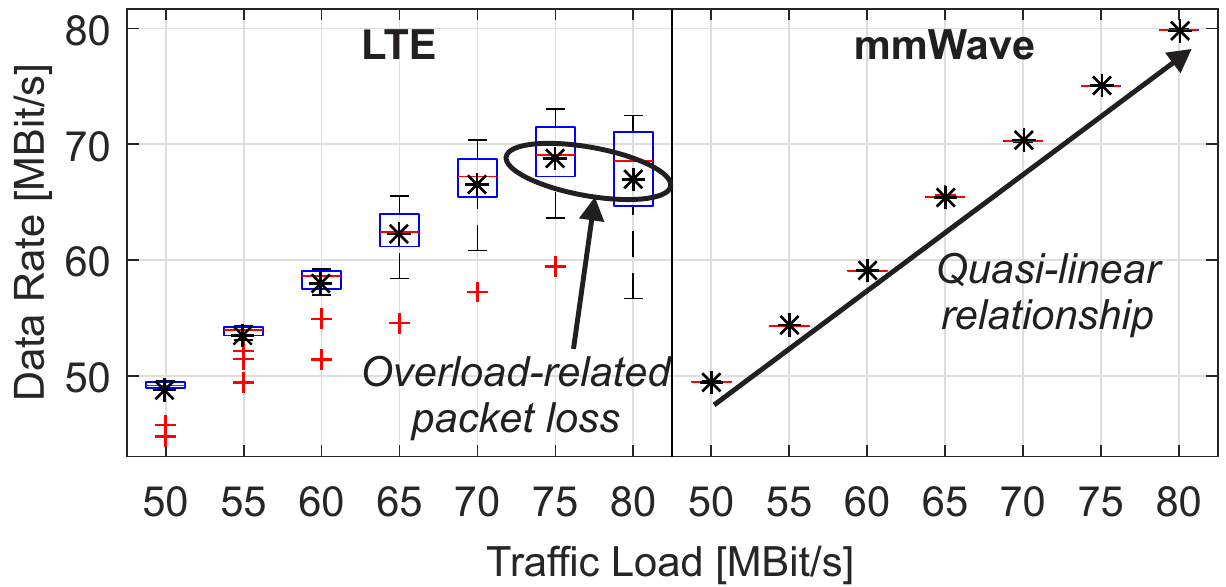}{0.44}{}
	\subfig{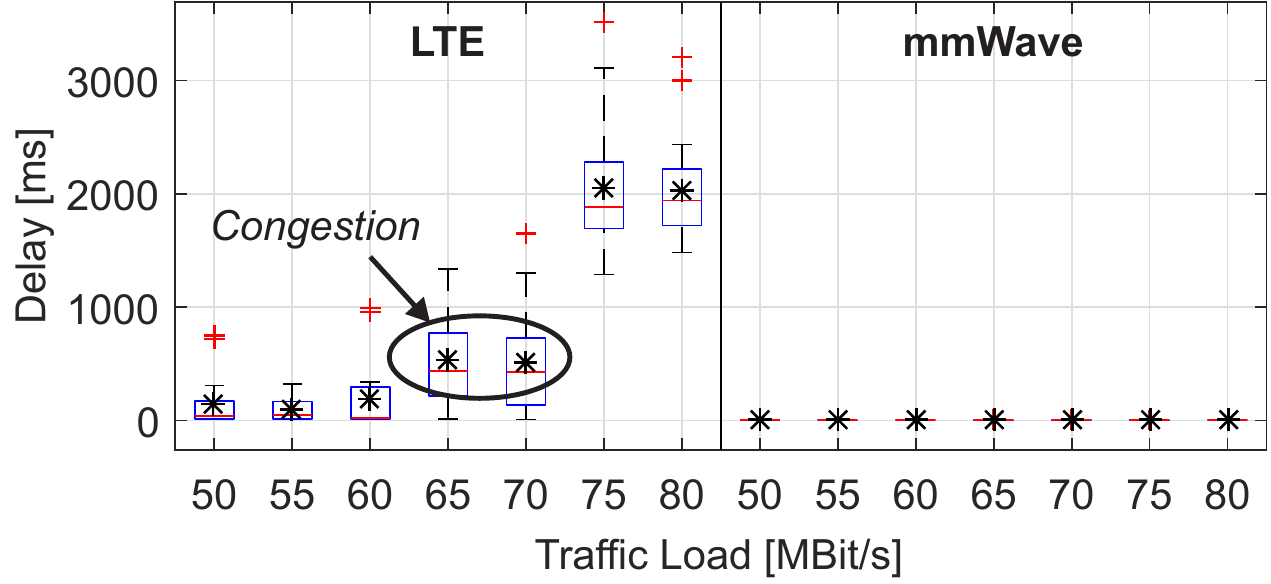}{0.44}{}

	\caption{Data rate and delay of \mmWave communications compared to a conventional \ac{LTE} link.}
	\label{fig:aerial_monitors_box}
\end{figure}

The overall results for data rate and delay are shown in Fig.~\ref{fig:aerial_monitors_box} for different amounts of traffic load.
While the \mmWave variant achieves a homogeneous performance for all simulated amounts of traffic load, different phases can be identified for \ac{LTE}:
%
% Low load
%
For lower traffic loads, the achieved data rate and delay behave similar for both radio link types.
%
% Medium load
%
The effects of a raising congestion level first manifest in the delay performance.
For medium traffic loads, the average delay as well as the delay variance are significantly increased due to the involved buffering effects. 
%
% High load
%
For high traffic loads, the network overload results in packet loss and a reduced data rate.
On the contrary, the large amount of available bandwidth allows the \mmWave technology to provide an approximately linear relationship between the offered traffic load and the resulting data rate with a low variance.
Additionally, the delay appears consistently low regardless of the offered traffic.

%
% Fig. PDR and delay of aerial sensors
%

\section{Conclusion} \label{sec:conclusion}

%
% Introduction
%
In this paper, we presented the open source \limosim framework which extends the \net ecosystem with support for joint simulation of hybrid ground-based and aerial communication networks based on the foundation of well-known analytical models for the low-level motion.
%
% Simulators
%
In contrast to existing approaches, \limosim couples mobility and communication simulation in an integrated way and in a single system process. This method enables direct (code-level) interactions between the entities of the different logical domains.
Based on two different case studies that focus on recent topics of vehicular networking, we have shown how \limosim can be integrated into \net-based hybrid vehicular network simulations.
%
% Future Work
%
Currently, we are investigating the integration of reinforcement learning-based mechanisms for mobility control and networking. 
Furthermore, we aim to bring together \limosim with \ac{DDNS} \cite{Sliwa/Wietfeld/2019b}. In addition, we will provide online visualization capabilities also for the communication processes.

\section*{Acknowledgment}

%\footnotesize
Part of the work on this paper has been supported by
the German Federal Ministry of Education and Research
(BMBF) in the project A-DRZ (13N14857) as well as by the
Deutsche Forschungsgemeinschaft (DFG) within the Collaborative Research Center SFB 876 ``Providing Information by Resource-Constrained Analysis'', project B4
% CC5G.NRW
and the Ministry of Economic Affairs, Innovation, Digitalisation and Energy of the state of North Rhine--Westphalia in the course of the Competence Center 5G.NRW under grant number 005--01903--0047.

\bibliographystyle{ACM-Reference-Format}
\bibliography{Bibliography}

\end{document}